\documentclass[conference]{IEEEtran}
\IEEEoverridecommandlockouts
\usepackage{cite}
\usepackage{amsmath,amssymb,amsfonts}
\usepackage{algorithmic}
\usepackage{graphicx}
\usepackage{subcaption}
\usepackage{textcomp}
\usepackage{xcolor}
\usepackage{multirow}
\usepackage{hyperref}
\usepackage{float}
\hypersetup{
    colorlinks=true,
    citecolor=black,
    linkcolor=black,
    filecolor=black,      
    urlcolor=black,
}
\def\BibTeX{{\rm B\kern-.05em{\sc i\kern-.025em b}\kern-.08em
    T\kern-.1667em\lower.7ex\hbox{E}\kern-.125emX}}

\begin{document}

\title{\huge \textbf{DualNet: Locate Then Detect Effective Payload \\ 
with Deep Attention Network}}

\author{\IEEEauthorblockN{Shiyi Yang\IEEEauthorrefmark{1}, Peilun Wu\IEEEauthorrefmark{2} and Hui Guo\IEEEauthorrefmark{3}}
\IEEEauthorblockA{School of Computer Science and Engineering, University of New South Wales, Sydney\IEEEauthorrefmark{1}\IEEEauthorrefmark{2}\IEEEauthorrefmark{3}\\
Innovation Institute, Sangfor Technologies Inc.\IEEEauthorrefmark{2}}
Email: \IEEEauthorrefmark{1}z5223292@cse.unsw.edu.au,
\IEEEauthorrefmark{2}wupeilun@sangfor.com.cn, \IEEEauthorrefmark{3}h.guo@unsw.edu.au
}

\maketitle

\begin{abstract}
Network intrusion detection (NID) is an essential defense strategy that is used to discover the trace of suspicious user behaviour in large-scale cyberspace, and machine learning (ML), due to its capability of automation and intelligence, has been gradually adopted as a mainstream hunting method in recent years.
However, traditional ML based network intrusion detection systems (NIDSs) are not effective to recognize unknown threats and their high detection rate often comes with the cost of high false alarms, which leads to the problem of alarm fatigue.

To address the above problems, in this paper, we propose a novel neural network based detection system, DualNet, which is constructed with a general feature extraction stage and a crucial feature learning stage. DualNet can rapidly reuse the spatial-temporal features in accordance with their importance to facilitate the entire learning process and simultaneously mitigate several optimization problems occurred in deep learning (DL). 
We evaluate the DualNet on two benchmark cyber attack datasets, NSL-KDD and UNSW-NB15.
Our experiment shows that DualNet outperforms classical ML based NIDSs and is more effective than existing DL methods for NID in terms of accuracy, detection rate and false alarm rate.
\end{abstract}



\section{Introduction}
With the ever-growing of network scale and complexity, cyber attacks are becoming more and more frequent, volatile and sophisticated, which imposes great threats to the massive networked society. 
The confidential information of the network users can be leaked; The integrity of data transferred over the network can be tampered; And the computing infrastructures connected to the network can be attacked.
Therefore, network intrusion detection system (NIDS) plays a pivotal role in offering the modern society a secure and reliable network communication environment.


Signature-based intrusion detection system (SIDS), due to its stability and dependability, is by far a typical type of NIDS that has been widely adopted in the commercial products.
SIDS relies on predefined attack signatures or patterns and can only detect known threats. By comparison, anomaly-based intrusion detection system (AIDS) exploits the capability of machine learning (ML) and uses the machine-learned rules and heuristics to identify deviations from normal network activities, making it possible for novel attacks to be detected.
However, ML-based NIDSs often achieve a high attack detection rate at the expense of many false alarms, which may cause the security team unnecessarily waste time on the fake threats and delay the responses to actual attacks.

Unlike many traditional ML algorithms that often require hand-designed features, DL can achieve much better generalization performance by self-learning its raw representations from the original dataset, which can potentially offer higher accuracy for network intrusion detection (NID). Though the existing DL methods do show such an advantage over the traditional ML approaches, the DL-based NIDS designs are not mature yet. The attack detection ability in the existing designs still need to be improved, and the false alarms are still not ignorable.

In this paper, we address these issues and propose a novel DL model, DualNet, for NID. DualNet can achieve a high learning accuracy and a high detection rate while keeping the false alarm rate and computational cost as low as possible. Our main contributions are summarized as follows:

\begin{itemize}

\item We introduce a novel neural network architecture, DualNet, that consists of two asynchronous stages: 1) a general feature extraction stage to maximally capture spatial and temporal features, and 2) a crucial feature learning stage to improve the detection efficiency by targeting important features for the final learning outcome.

\item We leverage a special learning method, densely connected learning, our work demonstrates that it exhibits no performance degradation and optimization difficulties in building deeper networks for NID.

\item We leverage the self-attention mechanism to effectively locate then detect the most valuable payloads from raw network packets in accordance with their scores of importance to improve the interpretability of DL for NID. 

\item We evaluate DualNet on two benchmark attack datasets, and compare our model with a set of existing ML and DL designs for NID. Our experiment results show that DualNet outperforms those existing designs.

\end{itemize}

A brief background of ML and DL for NID is provided in section \ref{bg_rw}. The design of densely connected learning and DualNet is presented in section \ref{sharknet}, and the evaluation of them is detailed in section \ref{e}. The paper is concluded in section \ref{conclusion}.

\section{Background and Related Work} \label{bg_rw}

In recent years, artificial intelligence (AI) based intrusion detection system has gained increasing popularity due to its ability of recognizing novel threats. The related NIDS designs can be divided into two categories: unsupervised learning based \cite{barlow1989unsupervised} and supervised learning based \cite{caruana2006empirical}.

Unsupervised learning builds a predictive profile based only on normal activities without the need to know any prior knowledge of attacks. Local Outlier Factor (LOF) \cite{kriegel2009loop} and K-means \cite{krishna1999genetic} are the typical design examples. 
These designs can reduce the cost required for data collection and corresponding labeling. It has been shown that they achieve a good performance in a controlled laboratory setting but are not so effective in a real network communication environment \cite{cs259d}.

Supervised learning, on the other hand, requires to learn the labelled datasets that cover both normal and malicious activities.
The approach shows a great potential on practical implementations \cite{suaboot2020taxonomy} and has been implemented in many designs: Some are based on classical machine learning (ML) algorithms and some are based on advanced deep learning (DL) methods. A brief review is given below.

\subsection{Classical Machine Learning Methods}  \label{bg_rw_a}

Among many classical ML methods \cite{buczak2015survey}, the kernel machines and ensemble classifiers are two effective strategies and are frequently applied to network intrusion detection (NID). 

Support Vector Machine (SVM) \cite{bao2009network} is a typical example of the kernel machine. It uses a kernel trick, such as radial basis function (RBF), to implicitly map the inputs to a high-dimensional feature space. However, SVM is not an ideal choice for heavy network traffic due to its high computation cost and moderate performance \cite{ahmad2018performance}.

Adaptive Boosting (AdaBoost) \cite{hu2013online} and Random Forest (RF) \cite{zhang2008random} are widely used ensemble classifiers. They incorporate multiple weak learners into a stronger learner to achieve a high accuracy that would not be possible from individual weak learners, and have powerful forces against overfitting. However, AdaBoost is sensitive to outliers and noisy data, and usually does not work well on imbalanced datasets. In contrast, RF can effectively handle imbalanced data. But because of its high computation complexity, it is slow in execution and not suitable for real-time intrusion detection \cite{ahmad2018performance}.

The traditional ML methods are often affected by so called `the curse of dimensionality' \cite{bengio2006curse}, the common bottleneck encountered during the design for performance optimization, which greatly limits the effectiveness of ML in learning the big data of increasing scale and complexity.
Another weakness of the ML based NIDS is that it often achieves high detection rate with the cost of high false alarms.

\subsection{Advanced Deep Learning Approaches} \label{bg_rw_b}

There are multiple DL approaches for network intrusion detection (NID), such as multilayer perceptron (MLP), convolutional neural networks (CNNs) and recurrent neural networks (RNNs). The DL based NIDS has a compelling capability to identify unknown attacks and has a high learning potential.

MLP \cite{ahmad2011intrusion} is an early kind of feed-forward artificial neural network (ANN) with multiple layers and non-linear activations. It adopts backpropagation \cite{hecht1992theory}, a supervised learning algorithm, for training.

CNNs \cite{xiao2019intrusion} are normally applied to capture spatial features from the learning dataset and produce feature maps as the outputs through convolution calculation. For one-dimensional security data, primitive CNN (ConvNet) \cite{vinayakumar2017applying} and depthwise separable CNN (DSC) \cite{lin2018using} are two effective detection methods in CNNs. Compared with ConvNet, DSC divides the whole convolution process into two simplified steps: depth-wise convolutions and point-wise convolutions, as such the number of multiplications and the number of trainable parameters can be reduced.

RNNs \cite{yin2017deep} are mainly used to extract temporal features from the network traffic records. Vanilla RNN fails to learn the long-term dependencies and suffers from the vanishing-gradient problem. To address these problems, long short-term memory (LSTM) \cite{althubiti2018lstm} has been proposed.
An advanced design, bidirectional LSTM (BiLSTM) \cite{goodfellow2016deep}, combines a forward LSTM with a backward LSTM, and it offers a high learning capability while at a considerable computational cost.
Gated recurrent unit (GRU) \cite{xu2018intrusion}, on the other hand, is a simplified LSTM with fewer number of gates and much lower trainable parameters.

In this paper, we propose DL model DualNet, which is a specially designed densely connected neural network (DenseNet) along with a self-attention mechanism. The model is presented in the next section. (The DenseNet was originally used for image recognition, and it's only for CNN, see \cite{huang2017densely}; The self-attention mechanism is mainly used for machine translation, more in \cite{vaswani2017attention}.)

\section{DualNet} \label{sharknet}

Our goal is to build a deep learning (DL) model that has a high detection capability (\textbf{model quality}) and is easy to train (\textbf{training efficiency}), and the trained model is small in size and fast in execution time (\textbf{model cost}). 

We consider that the model quality is closely related to the features extracted from the security data and how the extracted features are effectively used for the final prediction outcome. To this end, we propose a two-stage deep neural network architecture, DualNet: A \textbf{\textit{general feature extraction stage}} to maximally capture spatial-temporal features from the network traffic records; and a \textbf{\textit{crucial feature learning stage}} to focus more on important features to further improve the detection efficiency.

In terms of training efficiency and model cost, they are relevant to the number of trainable parameters, and a small trainable parameter number is desired. We, therefore, take this into account in our design. 

\begin{figure*}[t]
    \centering
    \includegraphics[width=.9\linewidth]{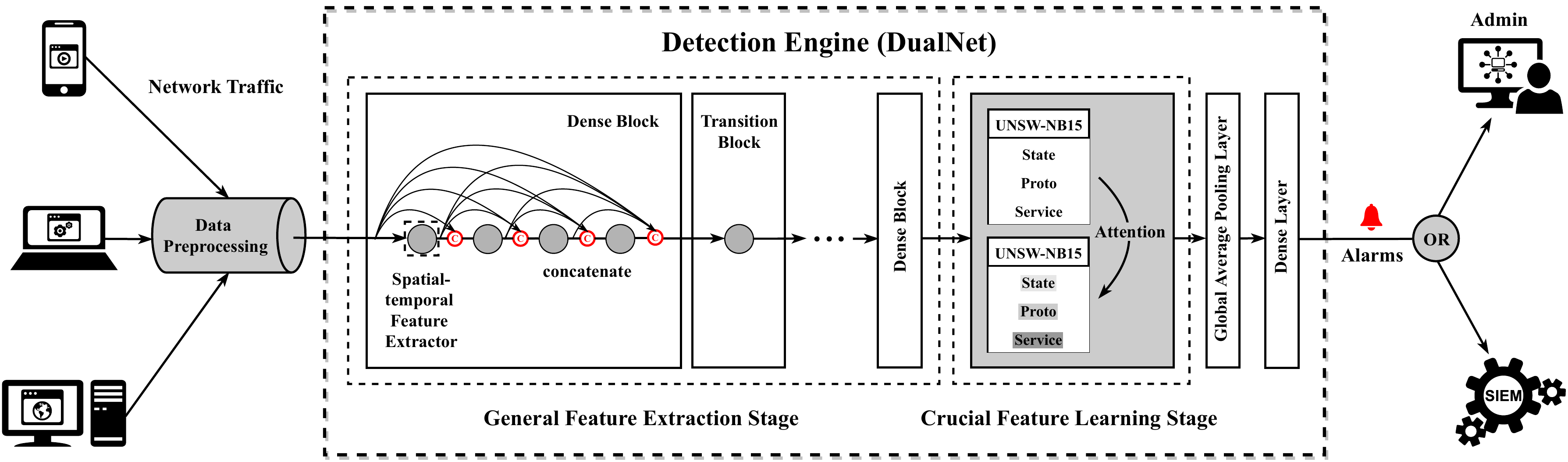}
    \caption{System Overview}
    \label{fig:systemoverview}
\end{figure*}

\begin{figure}[b]
\centering
    \includegraphics[width=.9\linewidth]{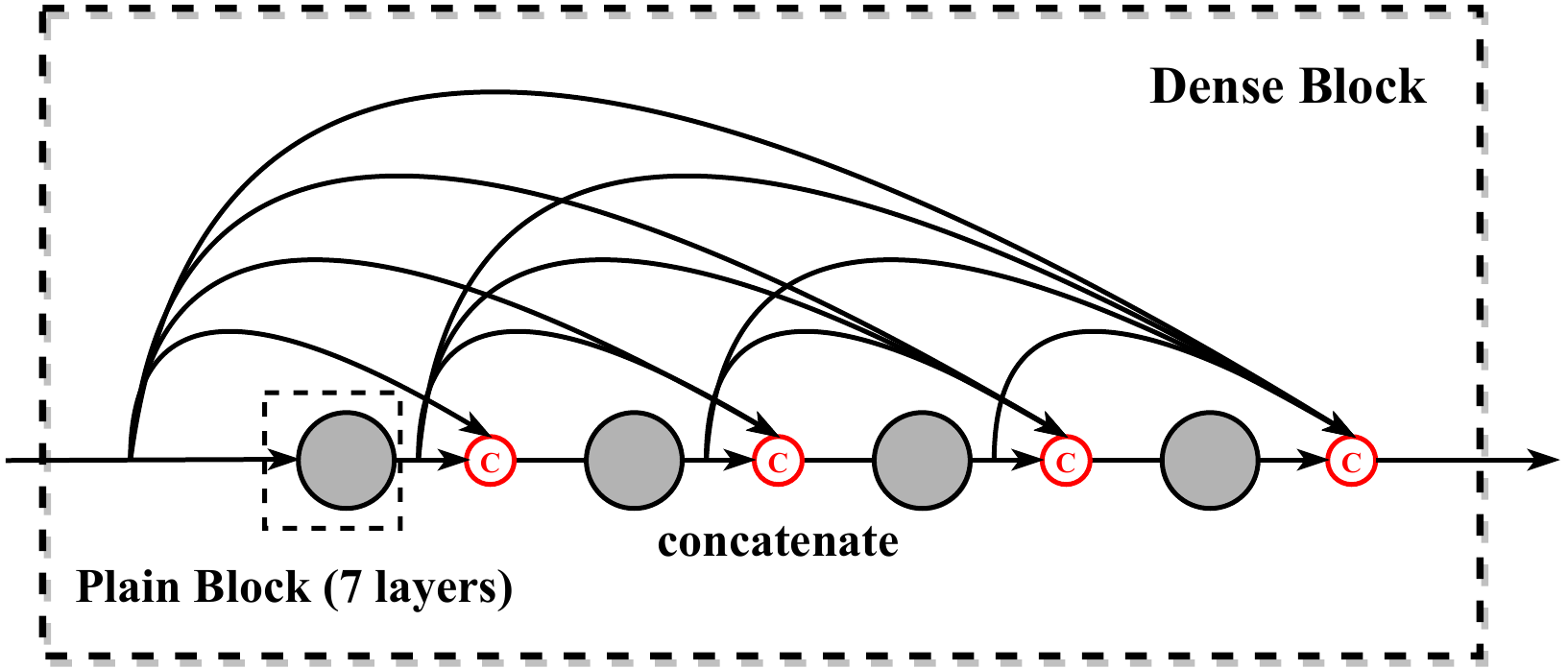}
    \caption{A dense block with a growth rate of k = 4}
    \label{fig:denseblk}
\end{figure}

An overview of our system is given in Fig.~\ref{fig:systemoverview}. The DualNet mainly performs two stages for attack recognition. The construction of two stages is elaborated in the next two sub sections.

\subsection{General Feature Extraction Stage} \label{3_a}
We consider that the multi-sourced security data have both spatial and temporal correlations. Hence, we present a special learning method, \textbf{densely connected learning}, which can maximally learn spatial-temporal features at various levels of abstraction from the input representations, and allow to build deeper neural network without performance degradation and optimization difficulties. The densely connected learning is to establish an interleaved arrangement pattern between specially designed blocks named \textbf{dense blocks} and particularly designed blocks called \textbf{transition blocks}, where the number of dense blocks is one more than the number of transition blocks, as shown in  Fig.~\ref{fig:systemoverview}. The design of dense blocks and transition blocks is detailed as below.

\smallskip

\subsubsection{Dense Block} \label{denseblk_structure}
Fig.~\ref{fig:denseblk} shows a dense block containing four specially designed basic blocks named \textbf{plain blocks}, where each plain block receives the concatenation of the output of all the preceding plain blocks and the input data through shortcut connections as its new inputs. We define a \textbf{growth rate k} to describe the number of plain blocks in each dense block.

\begin{figure}[t]
\centering
    \includegraphics[width=.9\linewidth]{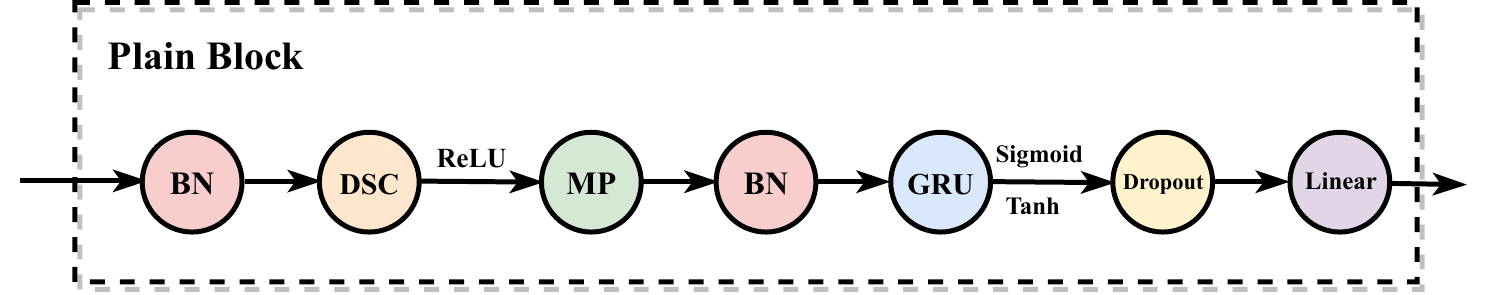}
    \caption{Plain Block (Spatial-temporal Feature Extractor)}
    \label{fig:plainblk}
\end{figure}

The plain block is a 7-layer (4 parameter layers) \textbf{spatial-temporal feature extractor}, as demonstrated in Fig.~\ref{fig:plainblk}. As discussed in section \ref{bg_rw_b}, the DSC and GRU require less trainable parameters. Hence, to efficiently leverage the feature extraction capability of both CNN and RNN for one-dimensional security data and reduce the potentially high computational cost of our densely connected learning, we combine them for building the plain blocks. Apart from DSC and GRU subnets, we also add five layers (including 2 parameter layers) to further enhance the learning ability:

\begin{itemize}
    \item Batch normalization (BN) \cite{ioffe2015batch} is applied to accelerate the training process and reduce final generalization error.
    \item Maxpooling (MP) layer is to provide basic translation invariance for the internal representations and decrease the computational cost.
    \item Dropout \cite{srivastava2014dropout}, a powerful regularization algorithm, which is used to counter overfitting trend. The dropout rate is adjusted to 0.4 here.
    \item Due to the randomness of neural network training, the results of each complete training process will be slightly different. Hence, a linear bridging strategy is appended to reduce the cost of retraining required to obtain the optimal model, and stabilize the learning process. Consequently, the model is not necessary to be retrained.
\end{itemize}

\begin{figure}[t]
     \centering
     \includegraphics[width=.9\linewidth]{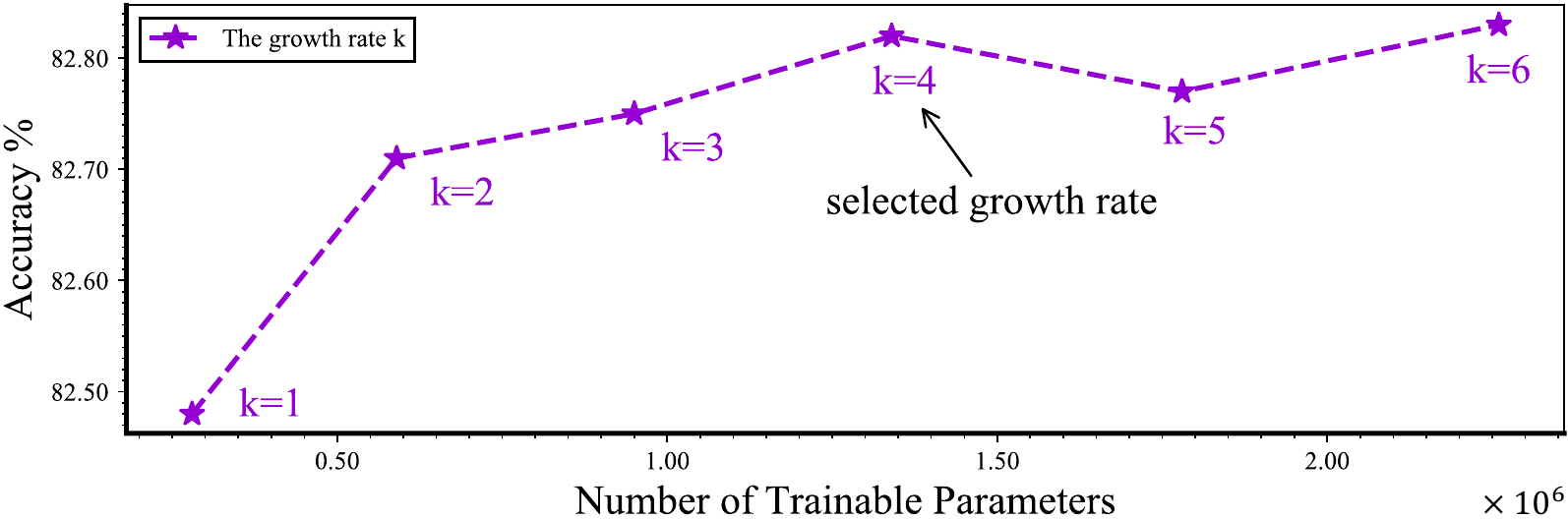}
     \caption{Testing accuracy and the number of trainable parameters of neural networks with only a dense block but under different growth rates k on UNSW-NB15 datasets}
     \label{fig:growth_rate}
 \end{figure}

To see how effective the growth rate k is for building dense blocks, we investigate the testing accuracy variation of neural networks with only a dense block but under different growth rates k on UNSW-NB15 \cite{moustafa2015unsw}. The experimental results are illustrated in Fig.~\ref{fig:growth_rate}. As shown in the figure, the accuracy initially improves with the growth rate. However, after k=4, further increasing the growth rate does little help to the accuracy, but just adds more trainable parameters. Therefore, we propose to fix the dense block with an optimal size with which the number of trainable parameter is small and the learning accuracy is high, such as k=4 for the given example.

The dense blocks encourage feature reuse and strengthen propagation of features and gradients within the network due to the dense connections. We can stack more dense blocks for a deeper neural network. 

\subsubsection{Transition Block} \label{transitionblk_structure}
The `curse of dimensionality' problem \cite{bengio2006curse} states that if the number of features (i.e. the dimensionality of feature space) of a neural network model increases rapidly, the prediction ability of the model will decrease significantly. The dense block with a growth rate k will increase the feature space dimensionality by $(k+1)$ times. Take the dense block shown in Fig.~\ref{fig:denseblk} as an example. The block has k equal to 4. It increases the dimensionality by 5 times, because five shortcut connections are concatenated as the outputs. Stacking one more block, the dimensionality will 25 times bigger. 
If $m$ such blocks are directly connected, the dimensionality would grow at the rate of $(k+1)^m$.

To mitigate the problem and continue to build deeper networks to fully learn the features at various levels of abstraction, we need to add a transition block between two dense blocks to reduce the dimensionality.

Since the DSC subnet has strong down-sampling capability, we use it for the dimensionality reduction. DSC favors the spacial features. To maintain both spacial and temporal features during the dimensionality reduction, we also add GRU subnet to the transition block. As a result, the transition block has the same structure as the plain block presented before. Inserting the block between dense blocks prevents the feature space grow, improving the generalization capability and robustness of the model and making the model easy to train.

\smallskip

In short, the first stage can be used to construct a very deep neural network with multiple dense blocks that are connected through transition blocks to extract general spatial-temporal features, as illustrated in  Fig.~\ref{fig:systemoverview}. To further improve the detection capability, we present the second stage to pay much attention to those features that are more important to the predicted results of the detection engine.

\subsection{Crucial Feature Learning Stage} \label{3_b}

We apply a self-attention mechanism \cite{vaswani2017attention} to focus more on the important features that should be considered as the most effective payloads to distinguish attack from normal behaviour.

In this stage, each feature will obtain an attention score, the higher its attention score, the more important it is and the more influence it has on the prediction of the detection engine.
The attention function can be described as mapping a query and a series of key-value pairs to an output that is specified as below.

\begin{equation}
Attention = softmax(Similarity(Q, K))V
\end{equation} where Q, K, V are the matrices of query, key, value respectively. The Similarity function performs dot-product calculation between the query and each key to obtain a weight, which is much faster and more space-efficient in practice \cite{vaswani2017attention}, that is, fewer trainable parameters are required. Finally, a softmax function is applied to normalize and assign these weights in conjunction with their corresponding values to obtain the final attention scores.

\begin{figure}[t]
    \centering
    \includegraphics[width=.9\linewidth]{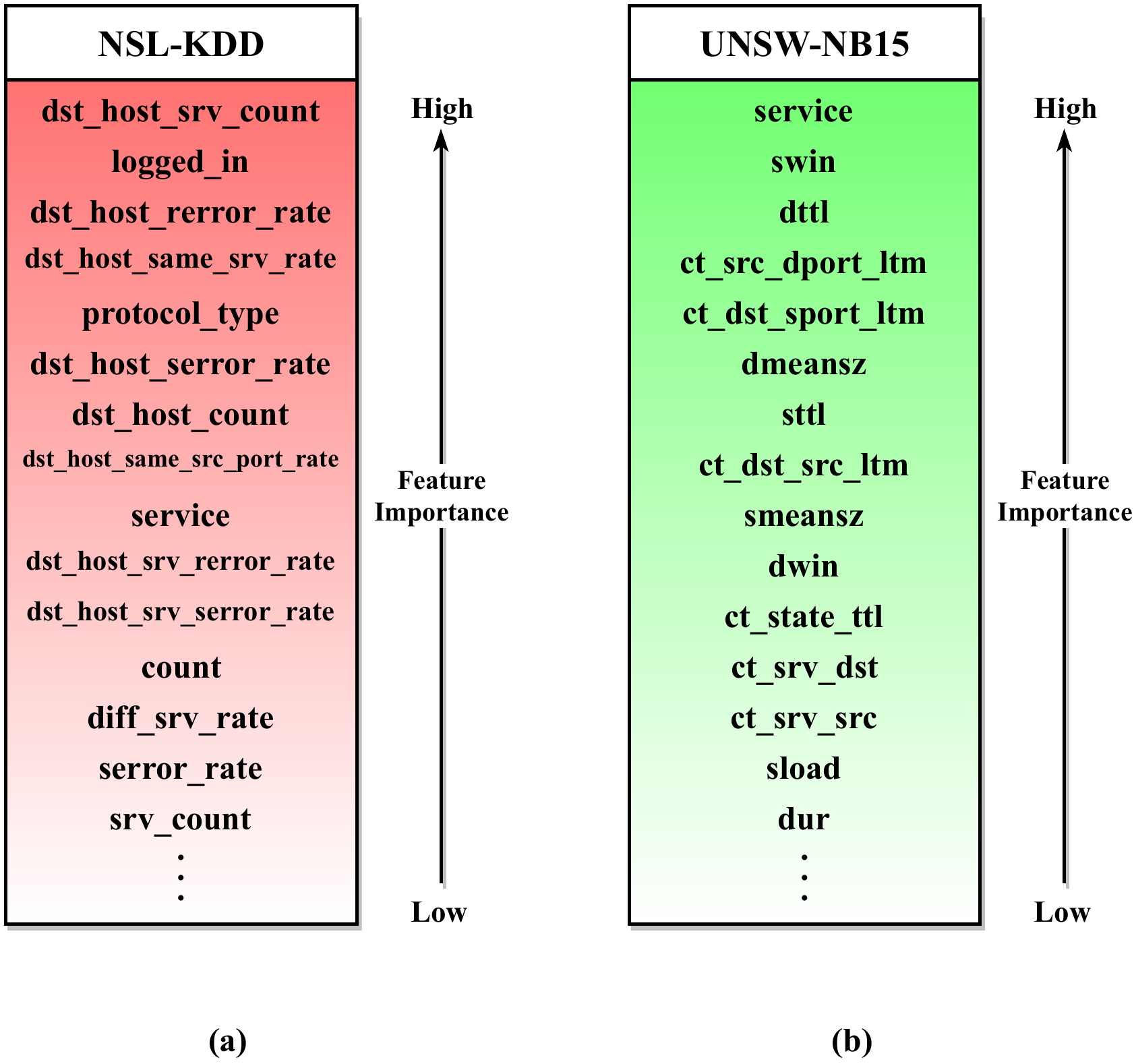}
    \caption{(a) The vital features are focused by self-attention mechanism on NSL-KDD datasets; (b) The vital features are focused by self-attention mechanism on UNSW-NB15 datasets}
    \label{fig:feature_importance}
\end{figure}

We conduct and visualize the attention score of each feature from the self-attention mechanism on NSL-KDD \cite{tavallaee2009detailed} datasets and UNSW-NB15 datasets \cite{moustafa2015unsw} respectively. Fig.~\ref{fig:feature_importance} shows the distribution of the top k most important features for the prediction on two datasets. Detailed result will be discussed in section \ref{model_performance}.

To sum up, the self-attention mechanism can enhance the interpretability of captured features and shrink the semantic gap between AI detectors and security analysts. Moreover, the mechanism can help security analysts obtain attention scores to pick out important features for the correlation analysis, thus further filtering false alarms to effectively identify real attacks and respond to attacks in time. Besides, by using the self-attention mechanism, our model can offer better capability to memorize long-term dependencies existed in the record to mitigate the gradient vanishing problem and performance degradation, thereby achieving higher accuracy.

\section{Evaluation} \label{e}
Our evaluation is based on a cloud AI platform configured with a Tesla K80 GPU and a total of 12 GB of RAM. The designs are written in Python building upon tensorflow backend with APIs of keras libraries and scikit-learn packages.

\renewcommand{\baselinestretch}{1.6}
\begin{table}[t]
\caption{\textsc{Ground Truth of UNSW-NB15}}
\begin{center}
\begin{tabular}[\linewidth]{c||c}
\hline
\textbf{Attack Category} & \textbf{Attack References (Description)}\\
\hline
\hline
Generic &  CVE-2005-0022, CVE-2006-3086, ...\\
\hline
Exploits & CVE-1999-0113, CVE-2000-0884, ...\\
\hline
Fuzzers & NULL (HTTP GET Request Invalid URI)\\
\hline
Reconnaissance & CVE-2001-1217, CVE-2002-0563, ...\\
\hline
DoS & CVE-2007-3734, CVE-2008-2001, ...\\
\hline
Shellcode & milw0rm-1308, milw0rm-1323, ...\\
\hline
Backdoors & CVE-2009-3548, CVE-2010-0557, ...\\
\hline
Analysis & NULL  (IP Protocol Scan)\\
\hline
Worms & CVE-2004-0362, CVE-2005-1921, ...\\
\hline
\end{tabular}
\label{tab:threat_model}
\end{center}
\end{table}
\renewcommand{\baselinestretch}{1}

\subsection{Datasets Selection}

The training and testing of designs is performed on two heterogeneous network intrusion detection datasets: NSL-KDD \cite{tavallaee2009detailed} and UNSW-NB15 \cite{moustafa2015unsw}. There are no duplicate network traffic records in both proposed datasets to ensure that the designs used in the evaluation do not favor more frequent records; and the designs with better detection rate for repetitive records will not bias their performance \cite{tavallaee2009detailed}, \cite{moustafa2016evaluation}.

These two cyber attacks datasets are composed of two classes, namely, normal and anomalous. In terms of traditional NSL-KDD benchmark, the abnormal includes 4 categories: Denial of Service (DoS), Probing (Probe), Remote to Local (R2L) and User to Root (U2R), where the attack samples are gathered based on a U.S. air force network environment. For modern UNSW-NB15 benchmark, there are 9 contemporary synthesized attack activities: Generic, Exploits, Fuzzers, Reconnaissance, DoS, Shellcode, Backdoors, Analysis and Worms, which are collected from Common Vulnerabilities and Exposures\footnote{CVE: https://cve.mitre.org/}, Symantec\footnote{BID: https://www.securityfocus.com}, Microsoft Security Bulletin\footnote{MSD: https://docs.microsoft.com/en-us/security-updates/securitybulletins}. It is worth noting that each attack event is simulated from a real-world attack scenario with a specific attack reference, as listed in table \ref{tab:threat_model}. The actual attack references used for our evaluation is based on the table but not limited to it, where it is in the range from CVE-1999-0015 to CVE-2014-6271.

\subsection{Data Preprocessing} \label{d_p}

There are 148,516 and 257,673 data records from NSL-KDD (41 features) and UNSW-NB15 (42 features) respectively used in the evaluation. Before training and testing, we preprocess the network traffic records in three phases.

\subsubsection{\textbf{Nominal Conversion}}
Since categorical data cannot be fed into neural networks straightforward, textual notations such as `http' and `smtp' are required to be converted to numerical form. Hence, we apply one-hot encoding \cite{garcia2015data} to encode multi-class variables into dummy representations to evade the classifier to assume a natural priority in the interior of features, and expand the sparsity of the data to accelerate the training.

\subsubsection{\textbf{Random Shuffling}}
We randomly disrupts the order between the records to prevent the selectivity of gradient optimization direction from severely declining due to the limitation of data regularity, hence reducing the tendency of overfitting, and expediting the convergence rate.

\subsubsection{\textbf{Dimension Normalization}}
The value of features in different dimensions does not contribute equally to the procedure of model fitting, which may give undue emphasis to inputs of larger magnitude to eventually result in a bias. Thus, we use min-max normalization \cite{patro2015normalization} to reshape the features on a scale of 0 to 1 to maintain certain numerical comparability and improve the stability as well as speed of backpropagation.

\begin{figure}[t]
    \centering
    \includegraphics[width=.9\linewidth]{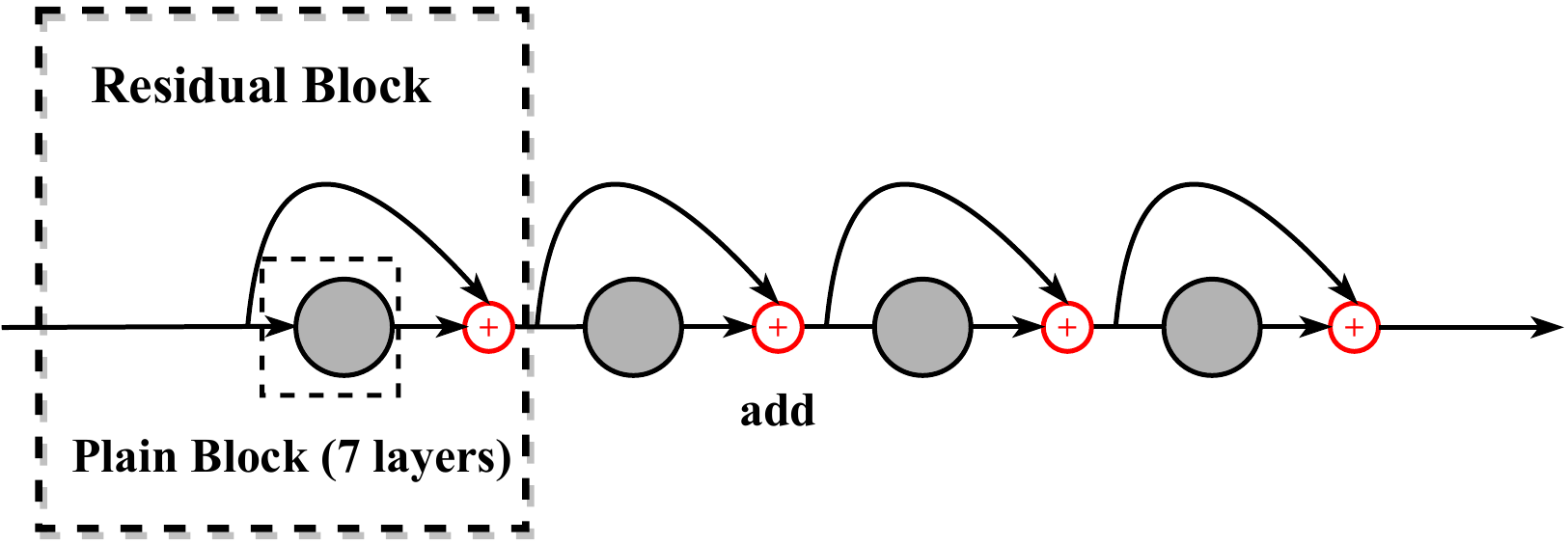}
    \caption{A ResNet with four residual blocks}
    \label{fig:resblk}
\end{figure}

\subsection{Training and Testing} \label{train_test}
To investigate the effectiveness of our densely connected learning in handling performance degradation problems and alleviating optimization difficulties, as well as its efficiency, and observe the effectiveness and efficiency of the self-attention mechanism for network intrusion detection (NID), we create three ResNets and three DenseNets in the same or simliar depths. The brief description is given below.

\textbf{ResNets}\textbf{.} Residual learning is originally used for image recognition and is only for CNN \cite{he2016deep}. Here, it is applied to a plain block to construct a special residual block: a ``skip" connection bipasses a plain block and is added to its output, as shown in Fig.~\ref{fig:resblk}. We name our ResNets $Residual-n$, where $n$ is the number of residual blocks. Each $Residual-n$ has $n$ residual blocks + one global average pooling layer + one dense layer: \textit{Residual-4 (31 layers including 19 parameter layers), Residual-8 (59 layers including 35 parameter layers), Residual-12 (87 layers including 51 parameter layers)}.

\textbf{DenseNets}\textbf{.} We apply our densely connected learning to establish the DenseNets. Similarly, we call our DenseNets $Dense-n$, where $n$ is the number of dense blocks with the growth rate k=4. Each $Dense-n$ has $n$ fix-sized dense blocks along with $(n-1)$ transition blocks in an interleaved arrangement pattern + one global average pooling layer + one dense layer: \textit{Dense-1 (31 layers including 19 parameter layers), Dense-2 (66 layers including 39 parameter layers), Dense-3 (101 layers including 59 parameter layers)}.

In essence, DualNet is the Dense-3 with a self-attention mechanism.

\subsubsection{Hyperparameter Settings}
To maintain a fair comparison for those networks, uniform hyperparameter settings are enforced for the training on two datasets separately. For all designs, the number of filters of convolution and the number of recurrent units are adjusted to be consistent with the number of features in each datasets, where NSL-KDD has 122 features and UNSW-NB15 has 196 features after the data preprocessing. Sparse categorical cross entropy loss function is used to calculate the errors, which sidesteps possible the memory constraints as a result of classification tasks with a large variety of labels. Adaptive moment estimation (Adam) algorithm is invoked as an optimizer, which computes individual adaptive learning rates for distinct parameters and generally leads to an outstanding performance of model especially for the sparse inputs \cite{ruder2016overview}. The learning rate is adjusted to 0.001 here.

\subsubsection{Stratified K-fold Cross Validation} We apply stratified k-fold cross validation to estimate the generalization ability of designs. The method splits the entire datasets into k groups by preserving the same proportion of each class in original records, where k-1 groups are combined for training and the remaining one is used for testing. Here, k is set to 10 to retain non-computational advantage of bias-variance trade-off \cite{james2013introduction}.

\begin{figure}[t]
    \centering
    \includegraphics[width=.9\linewidth]{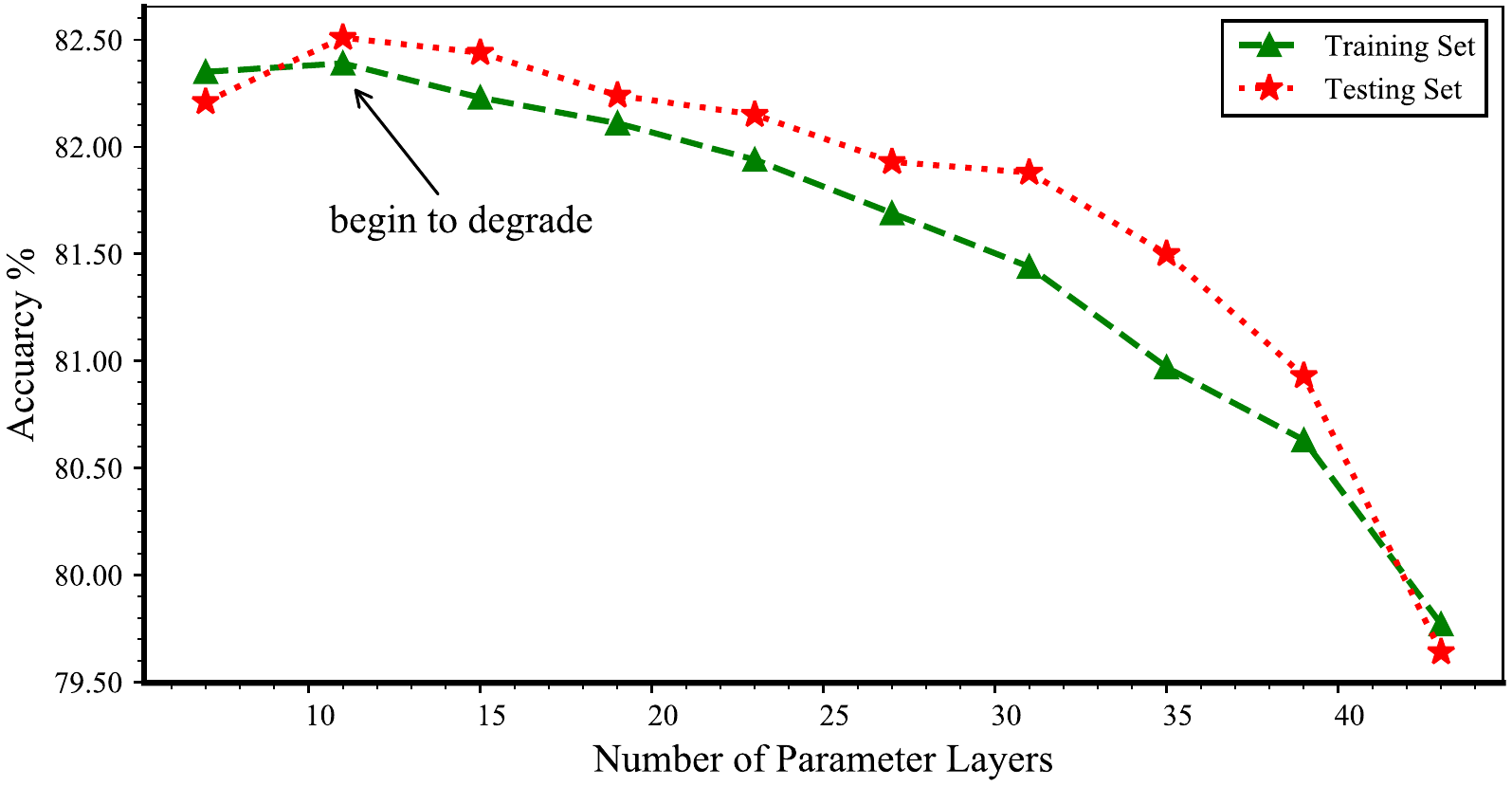}
    \caption{Performance degradation problem in building deeper networks for network intrusion detection on UNSW-NB15}
    \label{fig:performance_degradation}
\end{figure}

\subsubsection{Evaluation Metrics}
Three metrics are used to evaluate the performance of designs: Testing accuracy (ACC), detection rate (DR) and false alarm rate (FAR), as defined below.

\begin{equation}
ACC=\frac{Number\;of\;correct\;predictions}{Total\;number\;of\;predictions},
\end{equation}

\begin{equation}
DR=\frac{TP}{TP+FN},
\end{equation}

\begin{equation}
FAR=\frac{FP}{FP+TN},
\end{equation} where TP and TN are, respectively, the number of attacks and the number of normal network traffic accurately categorized; FP is the number of actual normal records misclassified as attacks, and FN is the number of attacks incorrectly classified as normal network traffic.

\subsection{DualNet Performance} \label{model_performance}

\begin{figure}[t]
     \centering
     \begin{subfigure}{\linewidth}
         \centering
         \includegraphics[width=.9\linewidth]{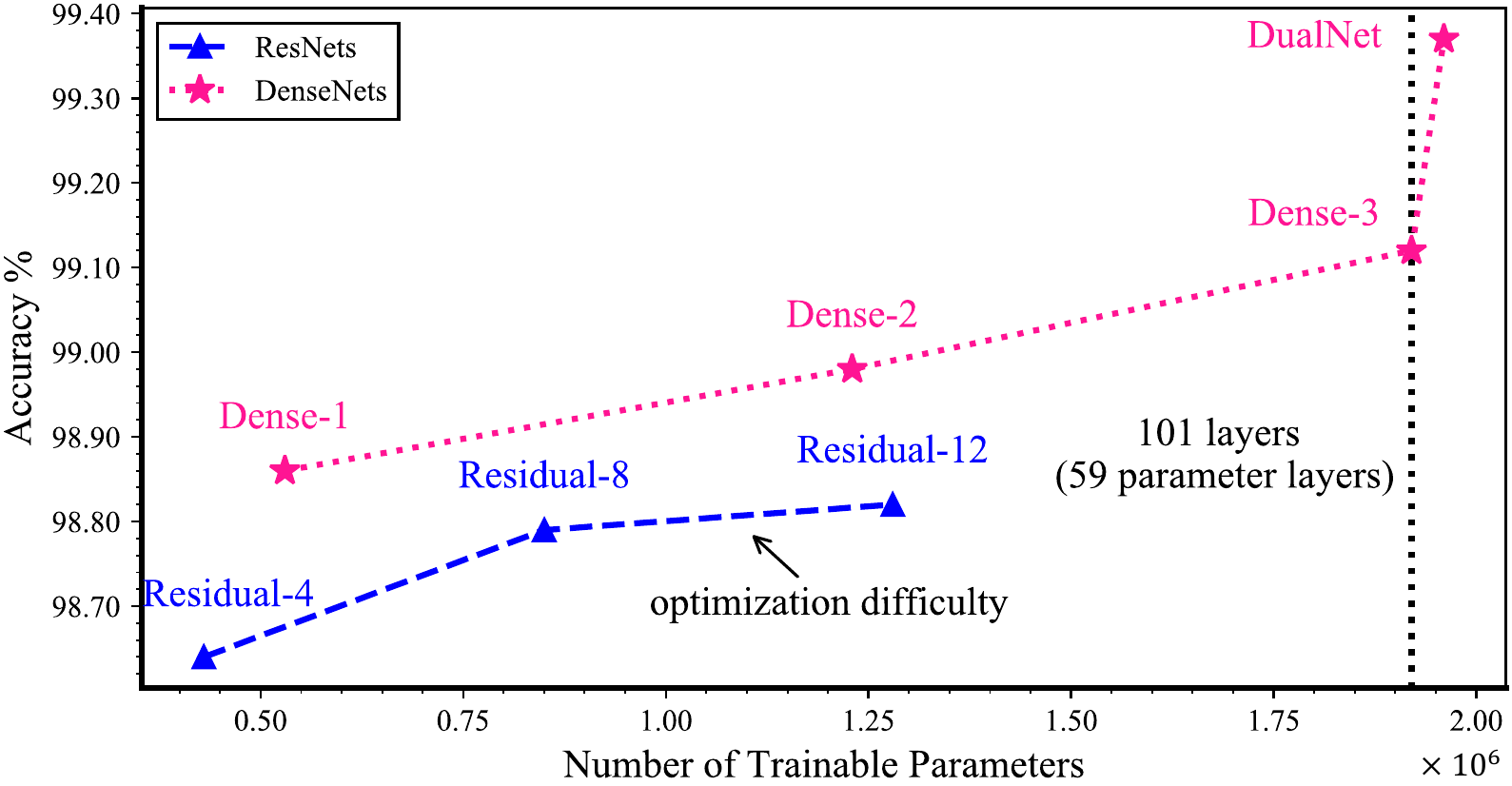}
         \caption{Evaluation metrics for seven designs on NSL-KDD}
     \end{subfigure}
     \hfill
     \begin{subfigure}{\linewidth}
         \centering
         \includegraphics[width=.9\linewidth]{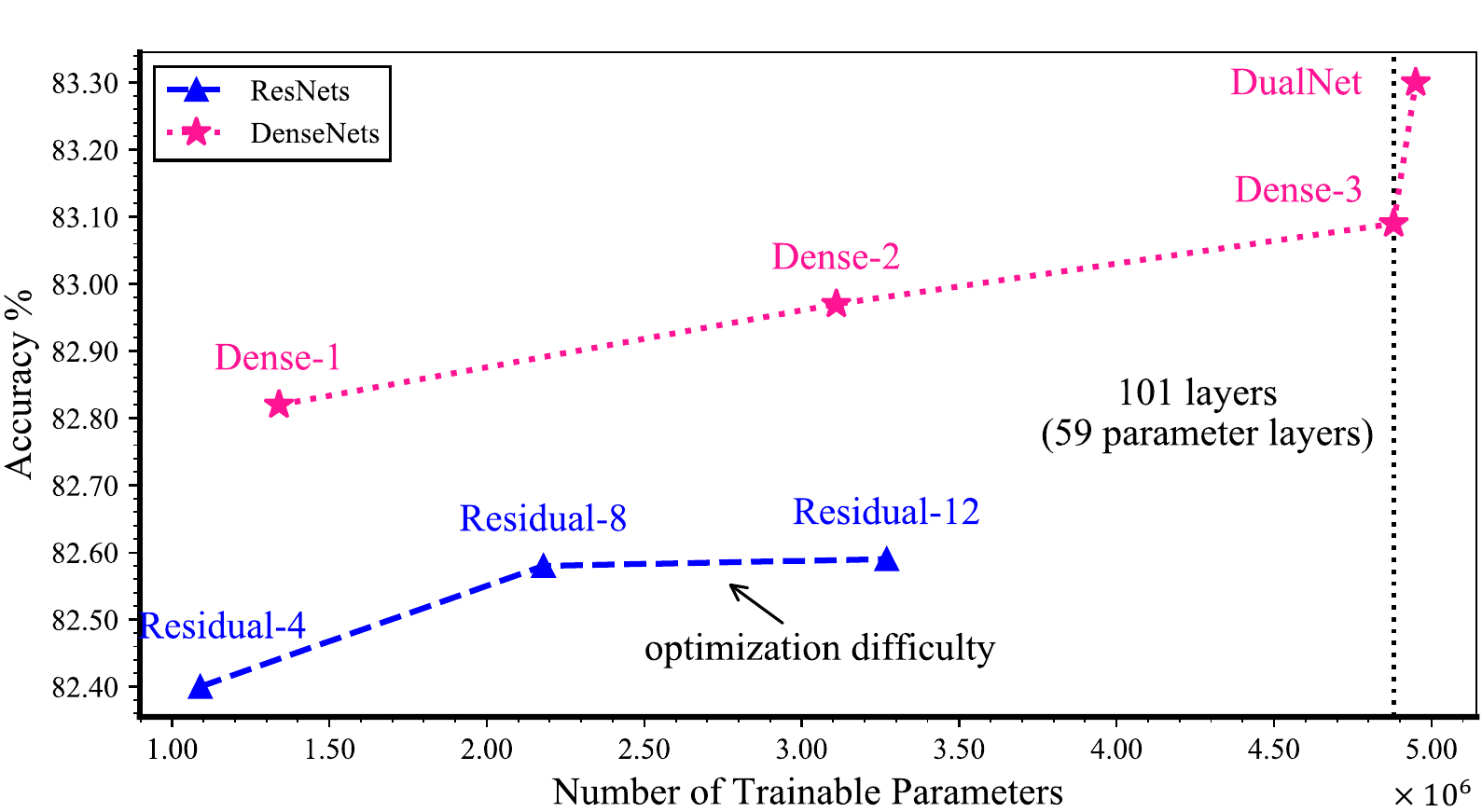}
         \caption{Evaluation metrics for seven designs on UNSW-NB15}
     \end{subfigure}
        \caption{Testing accuracy and the number of trainable parameters of seven designs on two datasets}
        \label{fig:seven_design}
\end{figure}

\renewcommand{\baselinestretch}{1.6}
\begin{table*}[htbp]
\caption{\textsc{Comparison Results of Methods on UNSW-NB15}}
\begin{center}
\begin{tabular}[\linewidth]{c|c||c|c|c|c|c|c|c}
\hline
\textbf{Type} & \textbf{Method} & \textbf{TP} & \textbf{FN} & \textbf{TN} & \textbf{FP} & \textbf{ACC \%}  & \textbf{DR \%} & \textbf{FAR \%}\\
\hline
\hline
\multirow{3}{*}{ML} & RF \cite{zhang2008random} & 10,156 & 6,311 & 8,716 & 584  & 61.02 & 61.67 & 6.28\\
\cline{2-9}
 & AdaBoost \cite{hu2013online} & 15,365 & 1,102 & 7,170 & 2,130  & 67.67 & 93.31 & 22.90\\
\cline{2-9}
 & SVM (RBF) \cite{bao2009network} & 13,463 & 3,004 & 8,639 & 661  & 72.86 & 81.76 & 7.11\\
\hline
\multirow{7}{*}{DL} & GRU \cite{xu2018intrusion} & 15,318 & 1,149 & 8,629 & 671  & 79.22 & 93.02 & 7.22\\
\cline{2-9}
& MLP \cite{ahmad2011intrusion} & 15,080 & 1,387 & 8,773 & 527 & 79.27 & 91.58 & 5.67\\
\cline{2-9}
& LSTM \cite{althubiti2018lstm} & 15,250 & 1,217 & 8,691 & 609 & 79.42 & 92.61 & 6.55\\
\cline{2-9}
& BiLSTM \cite{goodfellow2016deep} & 15,462 & 1,005 & 8,517 & 783 & 79.43 & 93.90 & 8.42\\
\cline{2-9}
& ConvNet \cite{vinayakumar2017applying} & 15,332 & 1,135 & 8,639 & 661 & 79.66 & 93.11 & 7.11\\
\cline{2-9}
& DSC \cite{lin2018using} & 15,306 & 1,161 & 8,774 & 526  & 80.16 & 92.95 & 5.66\\
\cline{2-9}
& \textbf{DualNet} & \textbf{15,555} & \textbf{912} & \textbf{8,816} & \textbf{484} & \textbf{83.30} & \textbf{94.46} & \textbf{5.20}\\
\hline
\end{tabular}
\label{tab:comparsion_results}
\end{center}
\end{table*}
\renewcommand{\baselinestretch}{1}

We first compare three DenseNets and three ResNets mentioned in section \ref{train_test} on two datasets, and then contrast DualNet to them. To in-depth evaluate the generalization performance of our model, we compare it with a series of existing ML and DL designs detailed in section \ref{bg_rw} on modern attacks datasets, UNSW-NB15. As a result, for network intrusion detection (NID), we have five observations as below.

\subsubsection{\textbf{Densely connected learning can handle performance degradation problem}} \label{d1}

We stack plain blocks from 1 to 10 to build the baseline comparison models to observe performance degradation problem in the construction of deeper neural networks for NID. Fig.~\ref{fig:performance_degradation} shows the training and testing accuracy of the network with respect to different number of parameter layers on UNSW-NB15 datasets. As can be seen from the figure, with the increase of network depth, the training and testing accuracy gets saturated at first and then declines rapidly as unexpected, namely, the performance gradually degrades. Fig.~\ref{fig:seven_design} illustrates the accuracy and the number of parameters of ResNets, DenseNets, and DualNet on two datasets. According to the figure, the learning accuracy improves when the network depth augments in the DenseNets on two datasets (Dense-2 outperforms Dense-1; Dense-3 outperforms Dense-2), which reflects our densely connected learning can effectively handle performance degradation problem in building deeper neural networks for NID.

\subsubsection{\textbf{Densely connected learning can alleviate optimization difficulties}} \label{d2} The optimization difficulty appears in the construction of deeper ResNets on two datasets, as shown in Fig.~\ref{fig:seven_design}, where Residual-12 is deeper than Residual-8 but they have very close accuracy. We consider that the ``add" operation in residual learning may hinder the transmission of information flow within the network \cite{huang2017densely}. Thereupon, we replace all the ``concatenate" connection modes in the DualNet with ``add" operation. Unexpectedly, the accuracy of using NSL-KDD datasets reduces from 99.37\% to 98.88\%, and it's down nearly 1\% on UNSW-NB15 datasets. Hence, the optimization difficulties in ResNets may be due to summation operations. By comparsion, the DenseNets exhibit no optimization difficulties, and the accuracy is greatly improved with the increase of depth, as shown in Fig.~\ref{fig:seven_design}. Therefore, our densely connected learning can alleviate optimization difficulties in constructing deeper neural networks for NID.

\subsubsection{\textbf{Densely connected learning is very efficient}} \label{d3} As can be seen from the Fig.~\ref{fig:seven_design}, DenseNets perform better than ResNets in the same or similar depths with achieving higher accuracy on two datasets. Incredibly, a shallower DenseNet can achieve better performance than a deep ResNet (Dense-1 outperforms Residual-8 and Residual-12, Dense-2 outperforms Residual-12), and it has lower trainable parameters. The results reflect the efficiency of densely connected learning for NID.

\subsubsection{\textbf{Self-attention mechanism is effective and efficient}} \label{d4}  
As displayed in Fig.~\ref{fig:seven_design}, compared to Dense-3, DualNet performs a sharp increase in accuracy while keeping a slight increase in trainable parameters on two datasets (99.37\% for NSL-KDD and 83.30\% for UNSW-NB15), which exhibits the effectiveness and efficiency of the self-attention mechanism for NID.

\renewcommand{\baselinestretch}{1.6}
\begin{table}[t]
\caption{\textsc{Evaluation Metrics of Using DualNet for Each Label on Two Datasets}}
\begin{center}
\begin{tabular}[\linewidth]{c|c||c|c|c}
\hline
\textbf{Datasets} & \textbf{Category} & \textbf{ACC \%}  & \textbf{DR \%} & \textbf{FAR \%}\\
\hline
\hline
\multirow{5}{*}{NSL-KDD} & Normal & 99.41 & 99.48 &  0.67\\
 \cline{2-5}
 & DoS & 99.92 & 99.96 & 0.10\\
  \cline{2-5}
 & Probe & 99.71 & 98.93 & 0.14\\
  \cline{2-5}
 & R2L & 99.38 & 92.23 & 0.27\\
  \cline{2-5}
 & U2R & 99.97 & 91.30 & 0.00\\
\hline
\hline
\multirow{10}{*}{UNSW-NB15} & Normal & 94.58 & 94.80 & 5.54\\
\cline{2-5}
 & Generic & 99.98 & 99.98 & 0.02\\
\cline{2-5}
 & Exploits & 98.98 & 97.69 & 0.41\\
 \cline{2-5}
 & Fuzzers & 89.82 & 66.49 & 4.61\\
 \cline{2-5}
 & Reconnaissance & 99.83 & 99.53 & 0.14\\
 \cline{2-5}
 & DoS & 99.80 & 88.11 & 0.01\\
 \cline{2-5}
 & Shellcode & 99.82 & 91.00 & 0.08 \\
 \cline{2-5}
 & Backdoors & 99.98 & 92.00 & 0.00\\
 \cline{2-5}
 & Analysis & 99.53 & 19.23 & 0.00\\
 \cline{2-5}
 & Worms & 100.00 & 100.00 & 0.00\\
\hline
\end{tabular}
\label{tab:each_attack}
\end{center}
\end{table}
\renewcommand{\baselinestretch}{1}

\subsubsection{\textbf{DualNet possesses an outstanding detection capability}} \label{comp_study}

Table \ref{tab:comparsion_results} illustrates TP, FN, TN, FP, ACC, DR and FAR of several existing ML and DL designs on UNSW-NB15 datasets. From the table, DualNet can identify more attacks (TP) with fewer omitted attacks (FN) and discover the maximum normal traffic (TN) with generating the minimum false alarms (FP). Moreover, our model significantly outperforms those designs with achieving higher ACC, higher DR and lower FAR. The comparsion results further demonstrate the effectiveness of DualNet for NID.

In addition to recognizing whether the network traffic record is normal or abnormal, DualNet can also identify a packet either as normal or as specific attacks. Table \ref{tab:each_attack} demonstrates ACC, DR and FAR of using our model for the normal and each attack on two datasets. From the table, DualNet exhibits an admirable ability to recognize normal network traffic and various specific attacks (DoS, Probe, R2L, U2R, Generic, Exploits, Reconnaissance, Shellcode, Backdoors and Worms) with a super ACC, a high DR along with a low FAR. Furthermore, the model maintains an acceptable level of capability for identifying Fuzzers attacks with marginally low DR at about 66.49\%. Nevertheless, there is a really low DR for Analysis attacks at approximately 19.23\%. The main reason is that about 65.54\% of Analysis are treated as the Exploits by the learner. It may be due to the overlap of important features or similar signatures between two attacks, and insufficient relevant training data (only approximately 1\% Analysis attacks records in UNSW-NB15 datasets), which confound the classifier.

All in all, DualNet performs a superior capability for precisely recognizing normal traffic and the abnormal one with achieving 99.33\% DR with 0.52\% FAR on NSL-KDD, and 94.46\% DR with 5.20\% FAR on UNSW-NB15.

\section{Conclusion} \label{conclusion}

In this paper, we propose a novel intrusion detection engine, DualNet, which is an extendable DenseNet with a self-attention mechanism. To capture both spacial and temporal features from the network traffic, we first build plain blocks with DSC and GRU subnets, based on which the dense blocks are created. In our design, the dense block offers a good trade off between learning accuracy and computer cost. To allow the neural networks grow deeper effectively, we interleave the dense blocks with transition blocks. Moreover, we investigate performance degradation in building deeper neural networks and optimization difficulties in constructing deeper ResNets for network intrusion detection (NID), and our densely connected learning can be applied to mitigate them effectively and efficiently. We also demonstrate the efficiency of the densely connected learning and the effectiveness and efficiency of the self-attention mechanism for NID.

Our experiments show that DualNet outperforms existing ML and DL designs for NID. Most importantly, its effectiveness on the  near real-world UNSW-NB15 dataset demonstrate its practical value to network security teams for traffic analysis and attack recognition.

\renewcommand{\baselinestretch}{1}
\bibliographystyle{./bibliography/IEEEtran}
\bibliography{./bibliography/IEEEabrv,./bibliography/IEEEexample}

\end{document}